\documentstyle[preprint,prb,aps]{revtex}
\begin{document}
\draft

\title{ {\small \tt Proc. of 38th M $\&$ MM Conference,
J. Appl. Phys. {\bf 75} (8), April (1994) }  \\
Phase transitions in the one-dimensional frustrated quantum
XY model and Josephson-junction ladders}

\author{Enzo Granato \\
Laborat\'orio Associado de Sensores e Materiais, \\
Instituto Nacional de Pesquisas Espaciais, \\
12225 S\~ao Jos\'e dos Campos, S\~ao Paulo, Brazil}

\maketitle

\begin{abstract}
A one-dimensional quantum version of the frustrated XY
(planar rotor) model is considered
which can be physically realized as a ladder of Josephson-junctions at half
a flux quantum per plaquette. This system undergoes a superconductor
to insulator transition at zero temperature as a function of charging
energy. The critical behavior  is
studied using a Monte Carlo transfer matrix applied to the
path-integral representation of the model and a finite-size-scaling
analysis of data on small system sizes. Depending on the ratio between
the interchain and intrachain couplings the system can have single or double
transitions which is consistent with the prediction that its critical
behavior should be described by the two-dimensional classical XY-Ising
model.
\end{abstract}

\newpage
The two-dimensional frustrated classical XY model has attracted
considerable attention recently \cite{Teitel,Berge,Gabay,%
GranatoK86,ThijssenK90,LeeKG,Jose,GranatoN}. It can be related to
Josephson-junction arrays in a magnetic field, where it is expected
to  describe
the finite-temperature superconductor to normal transition in arrays
with half a flux quantum per plaquette \cite{Physica}.
At low temperatures where capacitive
effects dominate, the array undergoes a superconductor to insulator
transition as a function of charging energy \cite{Geerligs,vdZant,%
GranatoK90,GranatoC}.
These charging effects arise from the
small capacitance of the grains making up the array and leads to strong
quantum fluctuations of the phase of the superconducting order
parameter. The critical behavior is now described by a two-dimensional
frustrated quantum XY model with a Hamiltonian \cite{GranatoK90}

\begin{equation}
H = -{{E_c}\over 2} \sum_r \left( {d \over { d \theta_r }} \right) ^2
- \sum_{<rr^\prime>} E_{rr^\prime} \cos ( \theta_r - \theta_{r^\prime})
\end{equation}
The first term in Eq. (1) describes quantum
fluctuations induced  by the charging energy  $E_c = 4 e^2/C$ of a
non-neutral superconducting grain located at site $r$, where $e$ is the
electronic charge and $C$ is the effective capacitance  of the grain.
The second term is the usual Josephson-junction coupling between
nearest-neighbor grains. $\theta_r$ represents the phase of the
superconducting order parameter  and the couplings $E_{rr^\prime}$
satisfy the Villain's 'odd rule' in which the number of negative bonds
in an elementary cell is odd \cite{Villain}. In a square lattice this
can be satisfied, for example, by ferromagnetic horizontal rows and
alternating ferromagnetic and antiferromagnetic columns of bonds.
This rule is a direct
consequence of
the constraint that, for the half flux case, the line integral of the
vector potential due to the applied magnetic field should be equal to $
\pi$ in units of the flux quantum.

In this work we consider a one-dimensional frustrated quantum XY
(1D FQXY) model \cite{Granato92} which can be regarded as the simplest 1D
version of the model (1) consisting just of a single column of
frustrated plaquettes as
indicated in Fig. 1. This can be physically realized as a periodic
Josephson-junction ladder at half a flux quantum per plaquette
\cite{Granato90,Kardar}.
In the classical limit ($E_c =
0$), the ground state of the 1D FQXY model
has a discrete $Z_2$ symmetry associated
with an antiferromagnetic pattern of plaquette chiralities $\chi_p =
\pm 1$, as indicated in Fig. 1,  measuring the two opposite directions
of the super-current
circulating in each plaquette. For small $E_c$, there is a gap for creation
of kinks in the antiferromagnetic pattern of $\chi_p$, and the ground state
has long-range chiral order.

Within a path-integral approach \cite{Granato90}, it can be shown that the
effective
action describing quantum fluctuations in the 1D FQXY model leads to two
coupled XY models in two (space-time) dimensions which is expected to have
a critical behavior in the universality class of the 2D
XY-Ising model \cite{GranatoK91} defined by the classical Hamiltonian
\begin{equation}
\beta H = - \sum_{<rr^\prime>} [ A(1+\sigma_r\sigma_{r^\prime}
\cos (\theta_r - \theta_{r^\prime}) + C \sigma_r \sigma_{r^\prime}]
\end{equation}
The phase diagram of this model consists of three branches, in the
ferromagnetic region. One of them corresponds to single transitions with
simultaneous loss of XY and Ising order. Further away from the branch
point, this line of single transitions becomes first order. The other
two lines correspond to separate XY and Ising transitions. The 1D FQXY
model is represented by a particular path through this phase diagram
which will depend on the ratio $E_x/E_y$ between the interchain and
intrachain couplings of the model. In this work we describe the results
of a finite-size scaling analysis of extensive calculations on
the 1D FQXY model \cite{Granato92,Granato93}
which shows that, in fact, the XY and
Ising-like excitations of the quantum model decouple for large
interchain couplings, giving rise to pure Ising model critical
behavior for the chirality order parameter. As a consequence, the
universality class of the superconductor-insulator transition in the
related Josephson-junction ladder should then depend on the ratio
between interchain and intrachain Josephson couplings.

To study the critical behavior of the 1D FQXY model, we find it
convenient to use an imaginary-time path-integral formulation of the
model \cite{ZinnJustin}.  In this formulation, the one-dimensional quantum
problem maps into a 2D classical statistical mechanics problem where the
ground state energy of the quantum model of finite size $L$ corresponds
to the reduced free energy per unit length of the classical model
defined on an infinite strip of width $L$ along the imaginary time
direction, where the time axis $\tau$ is discretized in slices $\Delta
\tau$. After scaling the time slices appropriately in order to get a
space-time isotropic model one obtains a classical partition function
where the parameter  $\alpha = (E_y/E_c)^{1/2}$ plays
the role of an inverse temperature in the 2D classical model.
The scaling behavior of the
energy gap for kink excitations (chiral domain walls) of the 1D FQXY
corresponds to the interface free energy of
an infinite strip in this classical model.  For large $\alpha$ (small
charging energy $E_c$), there is a gap for creation of kinks in the
antiferromagnetic pattern of $\chi_p$ and the ground state has
long-range chiral order. At some critical value of $\alpha$, chiral
order is destroyed by kink excitations, with an energy gap vanishing as
$| \alpha -\alpha_c | ^\nu$, which defines the correlation length
exponent $\nu$. Right at this critical point, the correlation function
decays as a power law $ <\chi_p \chi_{p^\prime} > = |p-p^\prime|
^{-\eta}$ with a critical exponent $\eta$.
The free energy  per unit length
$f(\alpha)$ of the Hamiltonian  on the infinite strip can be
obtained from the largest eigenvalue $\lambda_o$ of the
transfer matrix between different time slices as $f=- \ln \lambda_o$.
To obtain  $\lambda_o$, we used
a Monte Carlo transfer-matrix method \cite{Nightingale} which has
been shown to
lead to accurate estimates of the largest eigenvalue even for models
with continuous symmetry. The implementation of this method is similar
to the case of the 2D frustrated classical XY model
\cite{GranatoN} and further details
can be found in that work.

The interfacial free energy for domain walls  can be
obtained from the differences between the free energies for the
infinite strip with and without a wall. To obtain the critical
exponents and critical coupling we employ the finite-size scaling
$
\Delta F(\alpha, L) = A(L^{1/\nu} \delta \alpha)
$
where $A$ is a scaling function and $\delta \alpha= \alpha - \alpha_c$. In a
linear approximation for the argument of $A$, we have
\begin{equation}
\Delta F(\alpha, L) = a+ b  L^{1/\nu} \delta \alpha
\end{equation}
which can be used to determine the critical coupling $\alpha_c$ and the
exponent $\nu$ independently \cite{Granato92}.
The change from  an increasing trend with
$L$ to a decreasing trend provides and estimate of $\alpha_c$.
Once  the critical coupling
is known, the correlation function exponent $\eta$ can obtained from
the universal amplitude  $a$  in Eq. (3) through a result from
conformal invariance \cite{Cardy}, $a=\pi \eta$.
To estimate the correlation length exponent $\nu$ we
first obtain
the derivative $S= \partial \Delta F / \partial \alpha$  near
$\alpha_c$, then it can easily be seeing that a log-log plot of $S $ vs
$ L$ gives an estimate of $1/\nu$ without requiring a precise
determination of $\alpha_c$.  In Fig. 2 we show this kind
of plot for $E_x/E_y= 3$ from where we  get the
estimate $\nu= 1.05 (6)$. Of course, this is only valid in
the linear approximation of Eq. (3). To ensure that higher-order
terms can safely be neglected, the data for $\Delta F(\alpha,L)$
must be obtained in a sufficiently small range near $\alpha_c$.
We also checked, using a more general finite-size scaling analysis
\cite{GranatoN},
that allowing for higher-order terms gives results agreeing within the errors.
The results for the critical coupling $\alpha_c$
and critical exponents $\nu$
and $\eta$ for two different values of the ratio
$E_x/E_y= 1$ and $3$ are indicated in Table I.

For equal couplings
$E_x = E_y$, the results for the critical exponents $\eta$ and $\nu$
differ significantly from pure 2D Ising exponents ($\nu =1$ , $\eta=0.25$).
This result point to a single transition scenario. In fact, they
are consistent with a point along the line of single transitions in the
XY-Ising model \cite{GranatoK91}.
It is interesting to note that this result is also
consistent with similar calculations for the 2D frustrated classical
XY model \cite{GranatoN}.
In general, the critical behavior of a $d$ dimensional
quantum model is in the same universality class of the $d+1$ dimensional
classical version. However, the 1D FQXY model, apparently, is not the
Hamiltonian limit of the 2D classical model. Yet, their critical behavior
appears to be in the same universality class.

For the case of unequal couplings $E_x/E_y=3$, the results indicated in
Table I are in good agreement with pure 2D Ising values. From the
relation between the 1D FQXY model and the 2D XY-Ising model this then
implies that the XY and Ising-like excitations have decouple in this region
of parameters. To show that in fact one has two decoupled and, at the same
time, separated transitions we show in Fig. 3 the results for the helicity
modulus which measures the response of the system to an imposed twist.
The helicity modulus is  related to the
free-energy differences $\Delta F$ between strips with and without and
additional phase mismatch of $\pi$ along the strip and is given by
$\gamma = 2 \Delta F/\pi^2$ for large system sizes. If the model is
decoupled then  the transition should be in the universality class
of the 2D classical XY model, where one knows that the transition
is associated
with a universal jump of $2 /\pi$ in the helicity modulus \cite{Nelson}.
The
critical coupling can be estimated as the value of $\alpha$ at which
$\Delta F = \pi$ which gives $\alpha_c=1.29$.
This is to be compared with the critical coupling for the destruction
of chiral order in Table I, $\alpha_c = 1.16$. This clearly indicates
the transitions are well separated and thus one expects they are
decoupled.  We have also performed
less detailed  calculations at other values of the ratio $E_x/E_y$
from which we can estimate that the Ising and XY transitions merge into
a single transition roughly at $E_x/E_y \sim 2$. Since, the superconductor
to insulator transition is to be identified with
the loss of phase coherence \cite {Doniach} we reach the interesting result
that in the 1D FQXY, or alternatively, a Josephson-junction ladder, the
universality class of the superconductor-insulator transition depends on the
ratio between inter and intra-chain couplings.

\section{ Acknowledgments}

This work has been supported in part by Funda\c c\~ao
de Amparo \`a Pesquisa do Estado de S\~ao Paulo (FAPESP, Proc. no.
92/0963-5) and Conselho Nacional de Desenvolvimento Cient\'ifico e
Tecnol\'ogico (CNPq).

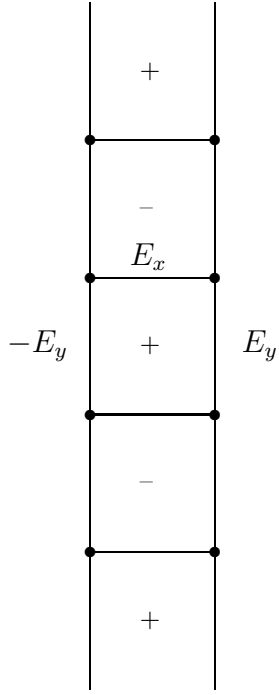
\begin{figure}
\caption{ Schematic representation of the one-dimensional frustrated
quantum XY
model with inter ($ E_x$) and intra-chain ($\pm E_y$) coupling constants.
The antiferromagnetic ordering of chiralities $\chi_p = \pm 1$ is also
indicated. }
\end{figure}
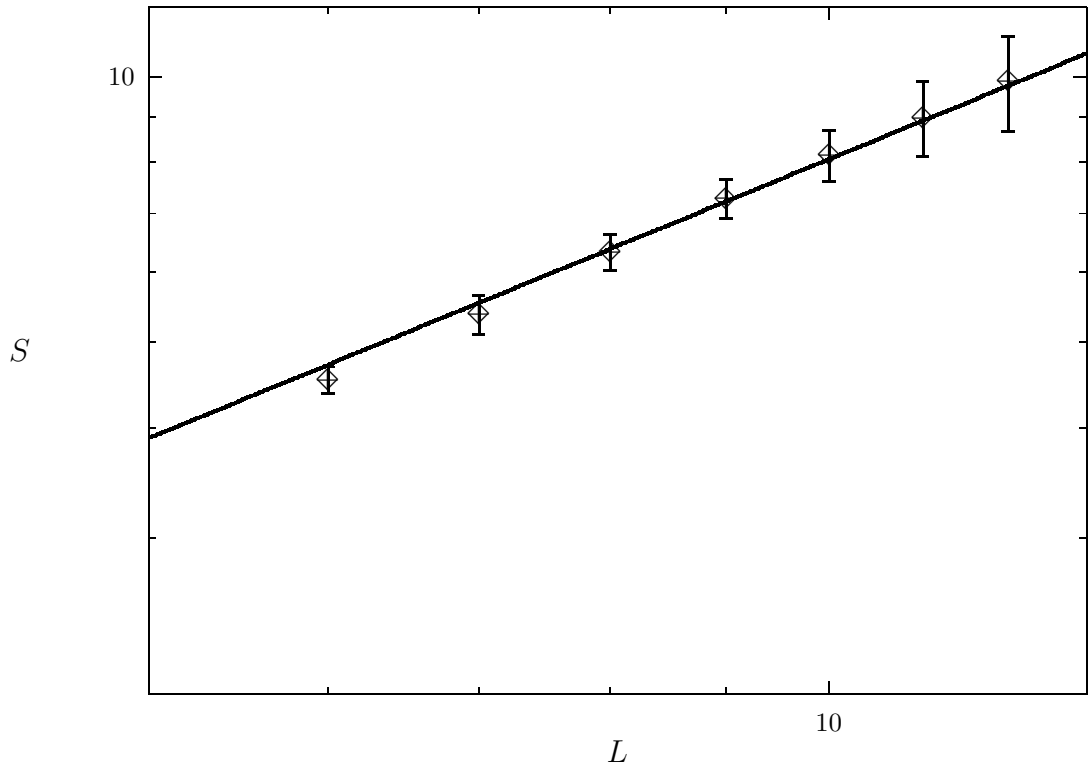
\begin{figure}
\caption{ $ S = \partial \Delta F(\alpha, L) / \partial \alpha$ evaluated near
the critical
coupling $\alpha_c$. The slope of the straight line gives an estimate of
$1/\nu$. }
\end{figure}
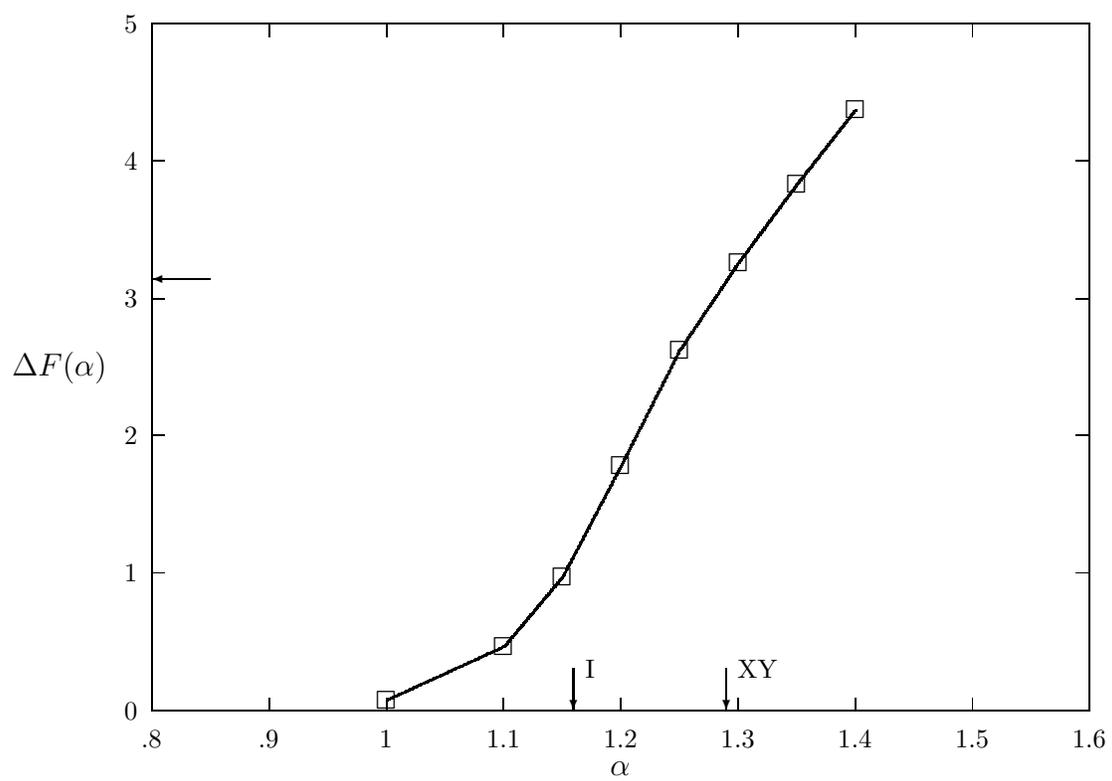
\begin{figure}
\caption{ Behavior of the interfacial free energy $\Delta F = L^2 \Delta f$
for a system of size $L=12$ resulting from an imposed  phase twist of $\pi$.
Vertical arrows indicate the locations of the Ising and XY transitions and
the horizontal arrow the value $\Delta F =\pi$
from where the XY transition is located.}
\end{figure}

\newpage
\begin{table}
\caption{ Critical  couplings ($\alpha_c=(E_y/E_c)^{1/2}$) and
critical exponents ($\nu$, $\eta$),
obtained from finite-size scaling analysis of
interfacial free energies for two values of the ratio
between interchain and intrachain couplings ($E_x/E_y$)  }

\vskip 1cm


\vfill Fig. 3
\end {center}

\end{document}